\documentclass[preprint,sort&compress]{elsarticle}

\usepackage{array}
\usepackage{longtable}
\usepackage{amssymb,amsmath}

\journal{J. of Geometry and Physics}

\newproof{proof}{Proof}
\newdefinition{remark}{Remark}

\newcommand {\ds}{\displaystyle}
\def\tr{\operatorname{trace}}
\def\sgrad{\operatorname{sgrad}}
\def\rank{\operatorname{rank}}

\usepackage{hyperref}

\begin{document}

\begin{frontmatter}
\author[fin]{P. E. Ryabov}
\ead{orelryabov@mail.ru}

\author[imash]{S.\,V.~Sokolov\corref{cor1}}
\ead{sokolovsv72@mail.ru}

\author[vags]{I.\,I.~Kharlamova}
\ead{irinah@vags.ru}

\address[fin]{Financial University, Leningradsky Avenue, 49, Moscow, 125993, Russia}

\address[imash]{Institute of Machines Science, Russian Academy of Sciences, 4 Maly Kharitonyevsky Per., Moscow, 101990, Russia}

\address[vags]{Russian Academy of National Economy and Public Administration,8 Gagarin Street, Volgograd, 400131, Russia}

\cortext[cor1]{Corresponding author}

\title{Explicit determination of certain periodic motions of a generalized two-field gyrostat\tnoteref{t1}}
\tnotetext[t1]{This work is partially supported by the grants of RFBR and Volgograd Region Authorities No. 14-01-00119, 15-41-02049, 16-01-00170 and 16-01-00809}

\begin{abstract}
The case of motion of a generalized two-field gyrostat found by V.\,V.~Sokolov and A.\,V.~Tsiganov
is known as a Liouville integrable Hamiltonian system with
three degrees of freedom. We find a set of points at
which the momentum map has rank 1. This set consists of special periodic motions
which correspond to the singular points of a bifurcation diagram on an iso-energetic surface.
For such motions the phase variables can be expressed in terms of algebraic functions
of a single auxiliary variable. These algebraic functions satisfy a differential equation integrable in elliptic
functions of time. It is shown that the corresponding points in the three-dimensional
space of the constants of the integrals belong to the intersection of two sheets of the
discriminant surface of the Lax curve.
\end{abstract}


\end{frontmatter}

\tableofcontents

\section{Intrtoduction}

The motion of a generalized two-field gyrostat is governed by the following
system of differential equations:
\begin{equation}\label{eq_1}
\begin{array}{l}
 \ds{\dot{\boldsymbol M}={\boldsymbol M}\times\frac{\partial H}{\partial{\boldsymbol
M}}+ {\boldsymbol\alpha}\times\frac{\partial H}{\partial{\boldsymbol\alpha}}+
{\boldsymbol\beta}\times\frac{\partial H}{\partial{\boldsymbol\beta}},}\\[5mm]
\ds{\dot{\boldsymbol \alpha}={\boldsymbol\alpha}\times\frac{\partial
H}{\partial{\boldsymbol M}},\quad \dot{\boldsymbol
\beta}={\boldsymbol\beta}\times\frac{\partial H}{\partial{\boldsymbol M}},}
\end{array}
\end{equation}
with Hamiltonian function \cite{SokTsi02}
\begin{equation}\label{eq_2}
\begin{array}{l}
H= M_1^2+M_2^2+2M_3^2+2\lambda M_3-2\varepsilon_2(\alpha_1+\beta_2)\\[3mm]
\qquad +2\varepsilon_1(M_2\alpha_3-M_3\alpha_2+M_3\beta_1-M_1\beta_3).
\end{array}
\end{equation}
Here ${\boldsymbol M}, {\boldsymbol\alpha}$ and ${\boldsymbol\beta}$ stand for the total angular momentum and the intensities of the two forces considered in the moving frame formed by the principal axes of inertia of the body. The gyrostatic momentum is directed along the axis of dynamic symmetry and its axial component is denoted by $\lambda$. The parameters $\varepsilon_1$ and $\varepsilon_2$ are called \textit{deformation parameters} since their zero values define important partial cases and establish relations with some previously known integrable cases.

Treating $\mathbb{R}^9=\{({\boldsymbol M}, {\boldsymbol\alpha}, {\boldsymbol\beta})\}$ as the Lie coalgebra $e(3,2)^*$ we obtain the Lie--Poisson bracket
\begin{equation}\label{eq_3}
\begin{array}{l}\{M_i,M_j\}=\varepsilon_{ijk}M_k, \quad \{M_i,\alpha_j\}=\varepsilon_{ijk}\alpha_k,\quad
\{M_i,\beta_j\}=\varepsilon_{ijk}\beta_k,\\[5mm]
\{\alpha_i,\alpha_j\}=0, \quad \{\alpha_i,\beta_j\}=0, \quad \{\beta_i,\beta_j\}=0, \\[5mm]
\varepsilon_{ijk}=\frac{1}{2}(i-j)(j-k)(k-i),\quad 1\leqslant i,j,k\leqslant 3.
\end{array}
\end{equation}
With respect to this bracket the system (\ref{eq_1}) can be represented in the Hamiltonian form
\begin{equation*}
\dot x=\{H,x\}
\end{equation*}
where $x \in \mathbb{R}^9$.

Note that the Casimir functions of the bracket (\ref{eq_3}) are ${\boldsymbol\alpha}^2$,
${\boldsymbol\alpha}\cdot{\boldsymbol\beta}$ and ${\boldsymbol\beta}^2$.
Therefore we define the phase space $\cal P$ of system \eqref{eq_1} as a common
level of these functions
\begin{equation*}
\boldsymbol\alpha^2=a^2,\quad \boldsymbol\beta^2=b^2,\quad
{\boldsymbol\alpha}\cdot{\boldsymbol\beta}=c, \quad (0<b<a, |c|<ab).
\end{equation*}
Using the parametric reduction invented by M.\,P.~Kharlamov \cite{Kh2005} we can assume that the parameter $c$ is zero. This simplifies calculations significantly.

In \cite{SokTsi02}, for the system \eqref{eq_1} with Hamiltonian function \eqref{eq_2},
V.\,V.\,Sokolov and A.\,V.\,Tsiganov gave a Lax representation with a spectral parameter
and thereby proved the Liouville complete integrability of this system. This Lax representation
generalizes the $L$-$A$ pair for the Kowalevski gyrostat in a double field found by A.\,G.\,Reyman and
M.\,A.\,Semenov-Tian-Shansky \cite{ReySem1987}.

For the Hamiltonian function \eqref{eq_2}, we represent the additional integrals
$K$ and $G$ as functions of two deformation parameters
$\varepsilon_1$ and $\varepsilon_2$ \cite{Rya13}:
\begin{equation*}
\begin{array}{l}
K=Z_1^2+Z_2^2-\lambda[(M_3+\lambda)(M_1^2+M_2^2)+2\varepsilon_2(\alpha_3M_1+\beta_3M_2)]\\[3mm]
\qquad+\lambda\varepsilon_1^2({\boldsymbol\alpha}^2+{\boldsymbol\beta}^2)M_3+2\lambda\varepsilon_1[\alpha_2M_1^2-\beta_1M_2^2-(\alpha_1-\beta_2)M_1M_2]
-2\lambda\varepsilon_1^2\omega_\gamma,\\[3mm]
G=\omega_\alpha^2+\omega_\beta^2+2(M_3+\lambda)\omega_\gamma-
2\varepsilon_2({\boldsymbol\alpha}^2\beta_2+{\boldsymbol\beta}^2\alpha_1)\\[3mm]
\qquad +2\varepsilon_1[{\boldsymbol\beta}^2(M_2\alpha_3-M_3\alpha_2)-
{\boldsymbol\alpha}^2(M_1\beta_3-M_3\beta_1)]\\[3mm]
\qquad+2({\boldsymbol\alpha}\cdot{\boldsymbol\beta})[\varepsilon_2(\alpha_2+\beta_1)+\varepsilon_1(\alpha_3M_1-\alpha_1M_3+\beta_2M_3-\beta_3M_2)].
\end{array}
\end{equation*}

Here we use the following notation:
\begin{equation*}
\begin{array}{l}
Z_1=\frac{1}{2}(M_1^2-M_2^2)+\varepsilon_2(\alpha_1-\beta_2)\\[3mm]
\qquad+\varepsilon_1[M_3(\alpha_2+\beta_1)-M_2\alpha_3-M_1\beta_3]+
\frac{1}{2}\varepsilon_1^2({\boldsymbol\beta}^2-{\boldsymbol\alpha}^2),\\[3mm]
Z_2=M_1M_2+\varepsilon_2(\alpha_2+\beta_1)-\varepsilon_1[M_3(\alpha_1-\beta_2)+\beta_3M_2-\alpha_3M_1]-\varepsilon_1^2(
{\boldsymbol\alpha}\cdot{\boldsymbol\beta}),\\[3mm]
\omega_\alpha=\alpha_1M_1+\alpha_2M_2+\alpha_3M_3,\quad
\omega_\beta=\beta_1M_1+\beta_2M_2+\beta_3M_3,\\[3mm]
\omega_\gamma=M_1(\alpha_2\beta_3-\beta_2\alpha_3)+M_2(\alpha_3\beta_1-\alpha_1\beta_3)+M_3(\alpha_1\beta_2-\alpha_2\beta_1).
\end{array}
\end{equation*}

In the special case where $\varepsilon_1=0$ and $\varepsilon_2=1$, we get the integrals of motion in the
problem of the Kowalevski gyrostat subjected to two homogeneous fields
\cite{ReySem1987,BobReySem1989}.

For the Lax pair due to Sokolov and Tsiganov \cite{SokTsi02}, the equation of the spectral curve ${\cal E}(z,\zeta)$ reads \cite{Rya13}
\begin{equation*}
{\cal E}(z,\zeta)\,:\, \, d_4\zeta^4+d_2\zeta^2+d_0=0,
\end{equation*}
where
\begin{equation*}
\begin{array}{l}
d_4=-z^4-\varepsilon_1^2({\boldsymbol\alpha}^2+{\boldsymbol\beta}^2)z^2-
\varepsilon_1^4[{\boldsymbol\alpha}^2{\boldsymbol\beta}^2-({\boldsymbol\alpha}\cdot{\boldsymbol\beta})^2],\\[3mm]
d_2=2z^6+[\varepsilon_1^2({\boldsymbol\alpha}^2+{\boldsymbol\beta}^2)-h-\lambda^2]z^4+
[\varepsilon_2^2({\boldsymbol\alpha}^2+{\boldsymbol\beta}^2)-
\varepsilon_1^2g]z^2\\[3mm]
\qquad+2\varepsilon_1^2\varepsilon_2^2[{\boldsymbol\alpha}^2{\boldsymbol\beta}^2-({\boldsymbol\alpha}\cdot{\boldsymbol\beta})^2],\\[3mm]
d_0=-z^8+hz^6+f_{\varepsilon_1,\varepsilon_2}z^4+\varepsilon_2^2gz^2-
\varepsilon_2^4[{\boldsymbol\alpha}^2{\boldsymbol\beta}^2-({\boldsymbol\alpha}\cdot{\boldsymbol\beta})^2].
\end{array}
\end{equation*}
The most complicated coefficient $f_{\varepsilon_1,\varepsilon_2}$ of $z^4$ in $d_0$ can be expressed in terms of the constants of the integrals  $h$, $k$, and $g$ as follows:
\begin{equation*}
f_{\varepsilon_1,\varepsilon_2}=\varepsilon_1^2g+k-\varepsilon_1^4({\boldsymbol\alpha}\cdot{\boldsymbol\beta})^2-
\frac{1}{4}[h^2+2\varepsilon_1^2({\boldsymbol\alpha}^2+{\boldsymbol\beta}^2)h+\varepsilon_1^4({\boldsymbol\alpha}^2-{\boldsymbol\beta}^2)^2]-
\varepsilon_2^2({\boldsymbol\alpha}^2+{\boldsymbol\beta}^2).
\end{equation*}

We define
\begin{equation*}
{\cal F}: {\cal P}\to {\mathbb R}^3
\end{equation*}
by ${\cal F}(x) =\{g = G(x), k = K(x), h = H(x)\}$. The mapping $\cal F$ is called a \textit{momentum mapping}.
By $\cal C$ we denote the set of all critical points of $\cal F$, i.e., the set of points $x$ such that $\rank d{\cal F}(x) < 3$. The set of critical values $\Sigma = {\cal F}({\cal C}) \subset {\mathbb R}^3$ is called the \textit{bifurcation diagram}. Normally, $\Sigma$ is a stratified 2-manifold.
In our case consider the bifurcation diagram $\Sigma\subset {\mathbb R}^3$ as a two dimensional
cell complex. Then the singular points form union of skeletons of dimensions
$0$ and $1$.
The determination of $1$-cells of $\Sigma$ is much more complicated. The corresponding values
of the first integrals are generated by periodic trajectories,
i.e., closed orbits for which $\rank d{\cal F}(x) = 1$. Let us call such a trajectory a \textit{special periodic motion}
(SPM) \cite{Kh2008}.
For the classical Kowalevski top in the gravity field all SPMs are permanent
rotations around the vertical axis. For a Kowalevski top in two constant fields ($\lambda = 0, \varepsilon_1=0$) the set
of SPMs, as shown in \cite{Kh2006}, consists of three families of pendulum motions pointed out in \cite{Kh2005}
for an arbitrary rigid body and the families of critical periodic motions of the Bogoyavlensky
case \cite{Bog1984}. These latter motions were first described in \cite{Zot2000} and explicitly integrated in \cite{Kh2006reg}.
Note that pendulum motions were first found by Yehia \cite{Yeh2001} with no conditions imposed on the
moments of inertia but under some special
restrictions on the  location of the centers of application of the fields.
For the integrable system of Kowalevski--Yehia SPMs were presented in \cite{KhPV1971},\cite{KhEIPV1969}.
Similar investigations in hydrodynamics were performed in \cite{Sok2013}, \cite{Bezg2014} and \cite{BorLeb1998}.

The aim of the present article is the construction of a certain class of periodic motions for the generalized two-field gyrostat. To do this we will study the singularities of the spectral curve associated with the Sokolov--Tsiganov $L$-$A$ pair.

\section{Discriminant surfaces and explicit integration of certain periodic motions}
The discriminant surface of the spectral curve ${\cal E}(z,\zeta)$ for Sokolov-Tsiganov $L-A$ pair was found in \cite{Rya15}
and consists of two surfaces $\Pi_1$ и $\Pi_2$:
\begin{equation*}
\Pi_1:\left\{
\begin{array}{l}
\ds{g(t)=\frac{ht^2-2t^3}{\varepsilon_2^2}+\frac{2\varepsilon_2^2(a^2b^2-c^2)}{t},}\\[3mm]
\ds{k(t)=3t^2-2ht+\frac{\varepsilon_1^2(2t^3-ht^2)}{\varepsilon_2^2}+\frac{\varepsilon_2^2(c^2-a^2b^2)(2\varepsilon_1^2t+\varepsilon_2^2)}{t^2}}\\[3mm]
\qquad \ds{+\frac{1}{4}\{h^2+2\varepsilon_1^2(a^2+b^2)h+\varepsilon_1^4[(a^2-b^2)^2+4c^2]+4\varepsilon_2^2(a^2+b^2)\}}.
\end{array}\right.
\end{equation*}
and
\begin{equation*}
\Pi_2:\left\{
\begin{array}{l}
\ds{g(s)=-\frac{\varepsilon_1^4\lambda^2[(a^2-b^2)^2+4c^2]}{s} -
\frac{\{\varepsilon_1^2[\varepsilon_1^2(a^2+b^2)+h+\lambda^2]+2\varepsilon_2^2\}}{8\varepsilon_1^8\lambda^2}s^2}\\[3mm]
\ds{\qquad+\frac{1}{16}\frac{s^3}{\lambda^2\varepsilon_1^8}+\frac{1}{2}\{\varepsilon_1^2[(a^2-b^2)^2+4c^2]+(a^2+b^2)(h+\lambda^2)\},}\\[5mm]
\ds{k(s)=\frac{\varepsilon_1^8\lambda^4[(a^2-b^2)^2+4c^2]}{s^2}+
 \frac{\{\varepsilon_1^2[\varepsilon_1^2(a^2+b^2)+h+\lambda^2]+2\varepsilon_2^2\}}{2\varepsilon_1^4}s}\\[3mm]
\ds{\qquad-\frac{3}{16}\frac{s^2}{\varepsilon_1^4}-\frac{1}{2}\lambda^2\bigl[2\varepsilon_1^2(a^2+b^2)+h+\frac{\lambda^2}{2}\bigr].}
\end{array}\right.
\end{equation*}

Consider a typical example of a discriminant surface on an iso-energetic level $h=const$.
\begin{figure}[ht]
\centering
\includegraphics[scale=0.2, clip]{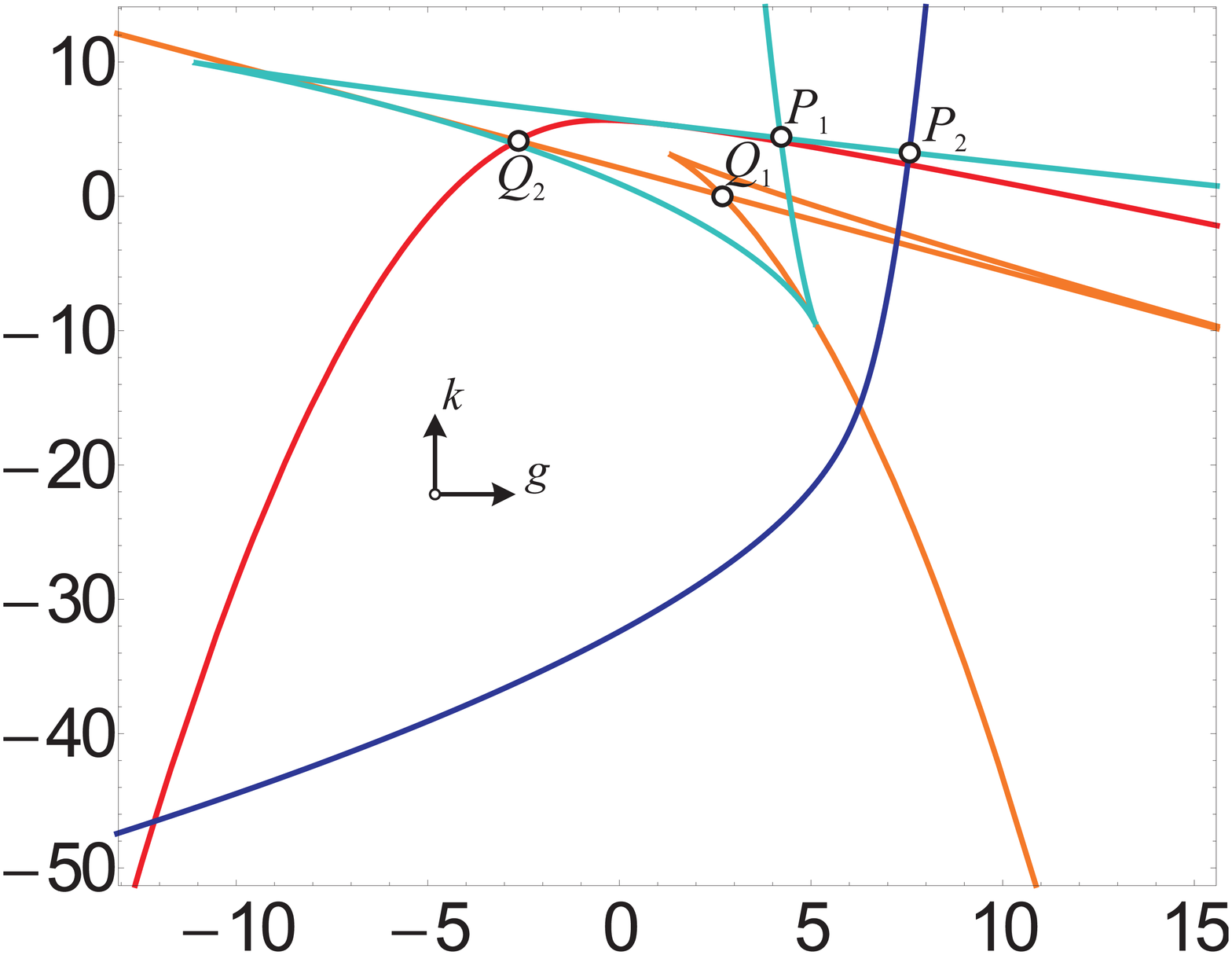}
\caption{Discriminant surfaces in the section $h=const$ for parameters $a = 1.;c = 0; \varepsilon_2 = 0.261; \varepsilon_1 = 0.851; b = 0.781; \lambda = 1.958; h = 3.41$.}\vskip 3mm
\includegraphics[scale=0.2, clip]{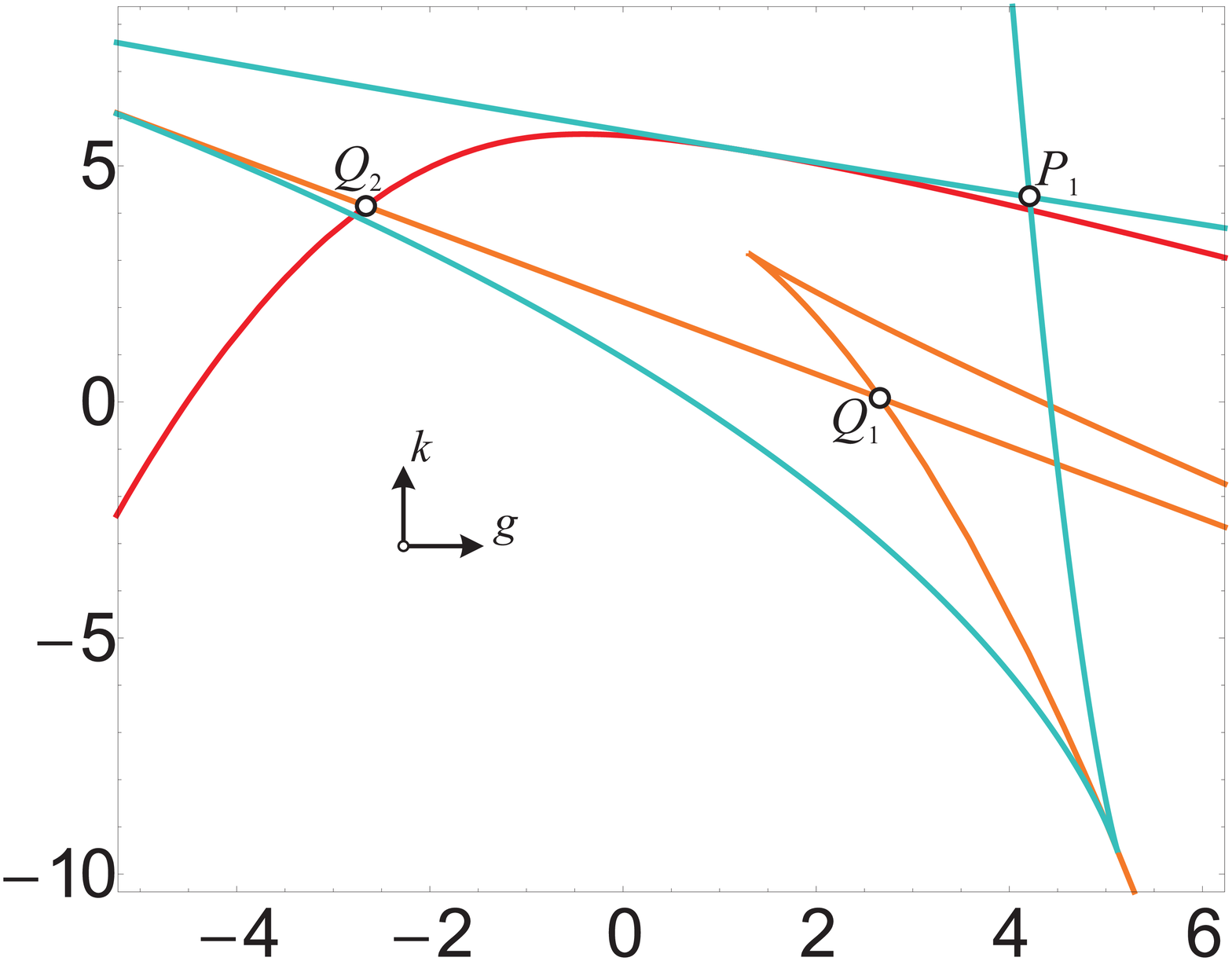}
\caption{Enlarged fragment.}\label{invfom}
\end{figure}

The points $P_1,P_2,Q_1$ and $Q_2$ are the self-intersection points of the discriminant leafs $\Pi_1$ and $\Pi_2$.
The coordinates of these points on the iso-energetic level $h=const$ are determined by the formulas
\begin{equation*}
P_1:\,\left\{\begin{array}{l}
\ds{g=b^2(h+\lambda^2)-\frac{1}{\varepsilon_1^2}(a^2-b^2)(\varepsilon_2^2+b^2\varepsilon_1^4),}\\
\ds{k=\frac{1}{4}[\varepsilon_1^2(a^2+b^2)+h]^2+\frac{\varepsilon_2^2}{\varepsilon_1^2}(h+\lambda^2)+
\frac{\varepsilon_2^4}{\varepsilon_1^4}+\varepsilon_2^2(a^2+b^2)-\varepsilon_1^2b^2\lambda^2,}
\end{array}\right.
\end{equation*}
\begin{equation*}
P_2:\,\left\{\begin{array}{l}
\ds{g=a^2(h+\lambda^2)+\frac{1}{\varepsilon_1^2}(a^2-b^2)(\varepsilon_2^2+a^2\varepsilon_1^4),}\\
\ds{k=\frac{1}{4}[\varepsilon_1^2(a^2+b^2)+h]^2+\frac{\varepsilon_2^2}{\varepsilon_1^2}(h+\lambda^2)+
\frac{\varepsilon_2^4}{\varepsilon_1^4}+\varepsilon_2^2(a^2+b^2)-\varepsilon_1^2a^2\lambda^2,}
\end{array}\right.
\end{equation*}
\begin{equation*}
Q_{1,2}:\,\left\{\begin{array}{l}
\ds{g=\pm abh,}\\
\ds{k=\frac{1}{4}(a\mp b)^2[2\varepsilon_1^2h+\varepsilon_1^4(a\pm b)^2+4\varepsilon_2^2].}
\end{array}\right.
\end{equation*}

Non-degenerate singularities of rank $1$ or special periodic motions are the preimages of the points mentioned above. How can one find them? By the definition of such a singularity \cite{BolFom} it
is required to find two function $g_1(x)$ and $g_2(x)$ for which at the points $x_0$ of rank $1$ the conditions $dg_1(x_0)=dg_2(x_0)=0$ are fulfilled.

The formulas for the coordinates of the points $P_k$ and $Q_k$ suggest that
\begin{equation*}
\begin{array}{l}
P_1: g_1=G-b^2H,\, g_2=K-\frac{1}{2}\left(h+\frac{1}{\varepsilon_1^2}[(a^2+b^2)\varepsilon_1^4+2\varepsilon_2^2]\right)H,\\[3mm]
P_2: g_1=G-a^2H,\, g_2=K-\frac{1}{2}\left(h+\frac{1}{\varepsilon_1^2}[(a^2+b^2)\varepsilon_1^4+2\varepsilon_2^2]\right)H,\\[3mm]
Q_{1,2}: g_1=G\mp abH,\, g_2=K-\frac{1}{2}(a\mp b)^2\varepsilon_1^2H.
\end{array}
\end{equation*}
At the singularities of rank $1$ (for which $P_1$ is the image) the following relation must hold:
\begin{equation}\label{eq_4}
\nabla G-b^2\nabla H+A\nabla (\alpha_1^2+\alpha_2^2+\alpha_3^2)+B\nabla (\beta_1^2+\beta_2^2+\beta_3^2) +C\nabla (\alpha_1\beta_1+\alpha_2\beta_2+\alpha_3\beta_3)=\mathbf{0}.
\end{equation}
Here $A,B$ and $C$ are undetermined multipliers.
Equations \eqref{eq_4} are equivalent to
\begin{equation*}
\sgrad G=b^2\sgrad H.
\end{equation*}
In this case the parametrization of special periodic motions looks like
\begin{equation}\label{eq_5}
\begin{array}{l}
\ds{M_1 =\frac{\sqrt{b^2\varepsilon_1^4-\varepsilon_2^2}[b(1+\xi^2)(1+\eta^2)
+2a\xi(1-\eta^2)]-2\lambda\varepsilon_1b(1+\eta^2)\xi}{\varepsilon_1b(1-\xi^2)(1+\eta^2)},}\\[5mm]
\ds{M_2=\varepsilon_2\frac{\sqrt{b^2\varepsilon_1^4-\varepsilon_2^2}[a(1+\xi^2)(1-\eta^2)+2b\xi(1+\eta^2)]-\lambda\varepsilon_1 b(1+\xi^2)(1+\eta^2)}{\varepsilon_1b\sqrt{b^2\varepsilon_1^4-\varepsilon_2^2}(1-\xi^2)(1+\eta^2)},}\\[5mm]
\ds{M_3=\frac{a\sqrt{b^2\varepsilon_1^4-\varepsilon_2^2}(1-\eta^2)-\lambda\varepsilon_1b(1+\eta^2)}{\varepsilon_1 b(1+\eta^2)},}\\[5mm]
\ds{\alpha_1=-2a\frac{\varepsilon_2(1-\eta^2)\xi-b\varepsilon_1^2(1-\xi^2)\eta}{b\varepsilon_1^2(1+\eta^2)(1+\xi^2)},\quad \alpha_2=a\frac{\sqrt{b^2\varepsilon_1^4-\varepsilon_2^2}(1-\eta^2)}{b\varepsilon_1^2(1+\eta^2)},}\\[5mm]
\ds{\alpha_3=-a\frac{\varepsilon_2(1-\eta^2)(1-\xi^2)+4b\varepsilon_1^2\eta\xi}{b\varepsilon_1^2(1+\eta^2)(1+\xi^2)},}\\[5mm]
\ds{\beta_1=\frac{2\sqrt{b^2\varepsilon_1^4-\varepsilon_2^2}\xi}{\varepsilon_1^2(1+\xi^2)},\quad
\beta_2=\frac{\varepsilon_2}{\varepsilon_1^2},\quad \beta_3=\frac{\sqrt{b^2\varepsilon_1^4-\varepsilon_2^2}(1-\xi^2)}{\varepsilon_1^2(1+\xi^2)}}.
\end{array}
\end{equation}
Here the auxiliary variables $\xi$ and $\eta$ satisfy the system of differential equations
\begin{equation*}
\begin{array}{l}
\ds{\dot\eta=-\varepsilon_1\frac{\sqrt{b^2\varepsilon_1^4-\varepsilon_2^2}[a(1+\xi^2)(1-\eta^2)+2b\xi(1+\eta^2)]-\lambda\varepsilon_1 b(1+\xi^2)(1+\eta^2)}{\sqrt{b^2\varepsilon_1^4-\varepsilon_2^2}(1-\xi^2)},}\\[5mm]
\ds{\dot\xi=-a\frac{\varepsilon_2(1-\eta^2)(1-\xi^2)+4\varepsilon_1^2b\xi\eta}{b\varepsilon_1(1+\eta^2)}.}
\end{array}
\end{equation*}
Using the equation $H=h$ one can eliminate $\eta$ and thus obtain a single equation in $\xi(t)$:
\begin{equation*}
(\dot\xi)^2=\frac{1}{\varepsilon_1^2\varepsilon_2^2}(a_4\xi^4+a_3\xi^3+a_2\xi^2+a_1\xi+a_0),
\end{equation*}
where
\begin{equation*}
\begin{array}{l}
a_4=-\varepsilon_2^2(-\varepsilon_2^2\varepsilon_1^2\lambda^2+4q^2\varepsilon_2^2+3q^4+2rq+\varepsilon_1^2q^2h),\\[3mm]
a_3=4\varepsilon_2^2\varepsilon_1\lambda(q\varepsilon_2^2+q^3+r),\\[3mm]
a_2=-4\varepsilon_1^2q^4h-8q^2\varepsilon_2^4+4r^2-8q^6-18q^4\varepsilon_2^2-2\varepsilon_1^2q^2\varepsilon_2^2h-2\varepsilon_1^2\varepsilon_2^4\lambda^2,\\[3mm]
a_1=4\varepsilon_2^2\varepsilon_1\lambda(q\varepsilon_2^2+q^3-r),\\[3mm]
a_0=-\varepsilon_2^2(-\varepsilon_2^2\varepsilon_1^2\lambda^2+4q^2\varepsilon_2^2-2rq+3q^4+\varepsilon_1^2q^2h).
\end{array}
\end{equation*}
Here the parameters $q$ and $r$ can be expressed in terms of $a$ and $b$ by formulas
\begin{equation*}
\begin{array}{l}
q=\sqrt{b^2\varepsilon_1^4-\varepsilon_2^2},\\[3mm]
r=\varepsilon_1^2\sqrt{(b^2\varepsilon_1^4-\varepsilon_2^2)[\varepsilon_1^2b^2(h+2\varepsilon_1^2b^2)+\varepsilon_2^2(a^2+b^2)]-\lambda^2b^2\varepsilon_1^2\varepsilon_2^2}.
\end{array}
\end{equation*}
These formulas can be used for construction of the characteristic polynomial which allows determination of the type of a singularity of rank $1$.

In the generic case the characteristic polynomial reads:
\begin{equation*}
\Delta(\mu)=\mu^4-p_2\mu^2-p_4,
\end{equation*}
where
\begin{equation*}
\begin{array}{l}
\ds{p_2=\frac{1}{2}\tr A_g^2,}\\[5mm]
\ds{p_4=\frac{1}{4}[\tr A_g^4-\frac{1}{2}(\tr A_g^2)^2].}
\end{array}
\end{equation*}
Here $A_g$ denotes a symplectic operator which is linearization of the vector field $\sgrad g$ at the points of solution \eqref{eq_5}.
Calculation of the coefficients $p_2$ and $p_4$ for the function $g_1$ at point $P_1$ yields
\begin{equation*}
\begin{array}{l}
\ds{p_2=-\frac{4}{\varepsilon_1^2}(a^2-b^2)
[a^2b^2\varepsilon_1^4-2\varepsilon_2^2a^2-3b^4\varepsilon_1^4-
\varepsilon_1^2b^2(h+\lambda^2)],}\\[5mm]
\ds{p_4=\frac{16}{\varepsilon_1^4}(a^2-b^2)^3[(b^2\varepsilon_1^4-\varepsilon_2^2)(\varepsilon_1^2b^2(h+2b^2)+\varepsilon_2^2(a^2+b^2))-\varepsilon_1^2\varepsilon_2^2\lambda^2b^2].}
\end{array}
\end{equation*}

Now let us describe periodic motions for the point $P_2$. Given that the skew-symmetric gradients are linearly dependent
\begin{equation*}
\sgrad G=a^2\sgrad H
\end{equation*}
one obtains the following parametrization
\begin{equation}\label{eq_6}
\begin{array}{l}
\ds{M_1=-\varepsilon_2\frac{\sqrt{a^2\varepsilon_1^4-\varepsilon_2^2}[b(1+\xi^2)(1-\eta^2)+2a\xi(1+\eta^2)]+\lambda\varepsilon_1 a(1+\xi^2)(1+\eta^2)}{\varepsilon_1a\sqrt{a^2\varepsilon_1^4-\varepsilon_2^2}(1-\xi^2)(1+\eta^2)},}\\[5mm]
\ds{M_2=-\frac{\sqrt{a^2\varepsilon_1^4-\varepsilon_2^2}[a(1+\xi^2)(1+\eta^2)
+2b\xi(1-\eta^2)]+2\lambda\varepsilon_1a(1+\eta^2)\xi}{\varepsilon_1a(1-\xi^2)(1+\eta^2)},}\\[5mm]
\ds{M_3=-\frac{b\sqrt{a^2\varepsilon_1^4-\varepsilon_2^2}(1-\eta^2)+\lambda\varepsilon_1a(1+\eta^2)}{\varepsilon_1 a(1+\eta^2)},}\\[5mm]
\ds{\alpha_1=\frac{\varepsilon_2}{\varepsilon_1^2},\quad
\alpha_2=\frac{2\sqrt{a^2\varepsilon_1^4-\varepsilon_2^2}\xi}{\varepsilon_1^2(1+\xi^2)},\quad
\alpha_3=\frac{\sqrt{a^2\varepsilon_1^4-\varepsilon_2^2}(1-\xi^2)}{\varepsilon_1^2(1+\xi^2)},}\\[5mm]
\ds{\beta_1=\frac{b\sqrt{a^2\varepsilon_1^4-\varepsilon_2^2}(1-\eta^2)}{a\varepsilon_1^2(1+\eta^2)},\quad
\beta_2=-2b\frac{\varepsilon_2(1-\eta^2)\xi-a\varepsilon_1^2(1-\xi^2)\eta}{a\varepsilon_1^2(1+\xi^2)(1+\eta^2)},}\\[5mm]
\ds{\beta_3=-b\frac{\varepsilon_2(1-\xi^2)(1-\eta^2)+4a\varepsilon_1^2\xi\eta}{a\varepsilon_1^2(1+\xi^2)(1+\eta^2)}.}
\end{array}
\end{equation}
Here the auxiliary variables $\xi$ and $\eta$ satisfy the equations
\begin{equation*}
\begin{array}{l}
\ds{\dot\eta=-\varepsilon_1\frac{\sqrt{a^2\varepsilon_1^4-\varepsilon_2^2}[b(1+\xi^2)(1-\eta^2)+2a\xi(1+\eta^2)]+\lambda\varepsilon_1 a(1+\xi^2)(1+\eta^2)}{\sqrt{a^2\varepsilon_1^4-\varepsilon_2^2}(1-\xi^2)},}\\[5mm]
\ds{\dot\xi=-b\frac{\varepsilon_2(1-\eta^2)(1-\xi^2)+4\varepsilon_1^2a\xi\eta}{a\varepsilon_1(1+\eta^2)}.}
\end{array}
\end{equation*}
Using the equation $H=h$ one can eliminate $\eta$ and thus obtain a single equation in $\xi(t)$
 \begin{equation*}
(\dot\xi)^2=\frac{1}{\varepsilon_1^2\varepsilon_2^2m^2}(b_4\xi^4+b_3\xi^3+b_2\xi^2+b_1\xi+b_0),
\end{equation*}
where
\begin{equation*}
\begin{array}{l}
b_4=-\varepsilon_2^2(4\varepsilon_2^2m^2-\varepsilon_1^2\lambda^2
\varepsilon_2^2+h\varepsilon_1^2m^2-2nm+3m^4),\\[3mm]
b_3=-4\varepsilon_2^2\varepsilon_1\lambda(\varepsilon_2^2m-n+m^3),\\[3mm]
b_2=-2\varepsilon_2^4\varepsilon_1^2\lambda^2-4\varepsilon_1^2hm^4-
8\varepsilon_2^4m^2-18\varepsilon_2^2m^4+4n^2-8m^6-
2\varepsilon_1^2h\varepsilon_2^2m^2,\\[3mm]
b_1=-4\varepsilon_2^2\varepsilon_1\lambda(\varepsilon_2^2m+n+m^3),\\[3mm]
b_0=-\varepsilon_2^2(4\varepsilon_2^2m^2-
\varepsilon_1^2\lambda^2\varepsilon_2^2+h\varepsilon_1^2m^2+3m^4+2nm).
\end{array}
\end{equation*}
Here the parameters $m$ and $n$ can be expressed in terms of $a$ and $b$ as follows
\begin{equation*}
\begin{array}{l}
m=\sqrt{a^2\varepsilon_1^4-\varepsilon_2^2},\\[3mm]
n=\varepsilon_1^2\sqrt{(a^2\varepsilon_1^4-\varepsilon_2^2)[\varepsilon_1^2a^2(h+2\varepsilon_1^2a^2)+\varepsilon_2^2(a^2+b^2)]-\lambda^2a^2\varepsilon_1^2\varepsilon_2^2}.
\end{array}
\end{equation*}
Finally the type of a singularity of rank $1$ which corresponds to the point $P_2$ can be determined with the help of the characteristic equation
\begin{equation*}
\Delta(\mu)=\mu^4-p_2^\prime\mu^2-p_4^\prime,
\end{equation*}
here
\begin{equation*}
\begin{array}{l}
\ds{p_2^\prime=-\frac{4}{\varepsilon_1^2}(a^2-b^2)
[a^2b^2\varepsilon_1^4-2\varepsilon_2^2b^2-3a^4\varepsilon_1^4-
\varepsilon_1^2a^2(h+\lambda^2)],}\\[5mm]
\ds{p_4^\prime=\frac{16}{\varepsilon_1^4}(a^2-b^2)^3[(a^2\varepsilon_1^4-\varepsilon_2^2)(\varepsilon_1^2a^2(h+2a^2)+\varepsilon_2^2(a^2+b^2))-\varepsilon_1^2\varepsilon_2^2\lambda^2a^2].}
\end{array}
\end{equation*}

At the point $Q_1$ we obtain a solution of the form
\begin{equation}\label{eq_7}
\begin{array}{l}
M_1=M_2=\alpha_3=\beta_3=0,\\[3mm]
\alpha_1=a\cos(\varphi),\alpha_2=a\sin(\varphi),\\[3mm]
\beta_1=-b\sin(\varphi),\beta_2=b\cos(\varphi).
\end{array}
\end{equation}
\begin{equation*}
\ddot{\varphi}=-8\varepsilon_2(a+b)\sin(\varphi)+4\varepsilon_1(a+b)
[\varepsilon_1(a+b)\sin(\varphi)-\lambda]\cos(\varphi).
\end{equation*}
Integrating gives
\begin{equation*}
(\dot\varphi)^2=4[\varepsilon_1^2(a+b)^2\sin^2\varphi+4\varepsilon_2(a+b)\cos\varphi-2\lambda\varepsilon_1(a+b)\sin\varphi+2h+\lambda^2].
\end{equation*}
Introduce a new variable $x(t)$:
\begin{equation*}
x=\tan \frac{\varphi}{2}.
\end{equation*}
Then
\begin{equation*}
\cos\varphi=\frac{1-x^2}{1+x^2},\quad\sin\varphi=\frac{2x}{1+x^2},\quad\dot\varphi=\frac{2\dot x}{1+x^2}
\end{equation*}
The differential equation in $x=x(t)$ looks like
\begin{equation*}
(\dot x)^2=c_4x^4+c_3x^3+c_2x^2+c_1x+c_0,
\end{equation*}
here
\begin{equation*}
\begin{array}{l}
c_4=2h+\lambda^2-4\varepsilon_2(a+b),\\[3mm]
c_3=-4\lambda\varepsilon_1(a+b),\\[3mm]
c_2=2[2\varepsilon_1^2(a+b)^2+2h+\lambda^2],\\[3mm]
c_1=-4\lambda\varepsilon_1(a+b),\\[3mm]
c_0=\lambda^2+2h+4\varepsilon_2(a+b).
\end{array}
\end{equation*}
Therefore $x(t)$ can be written in terms of elliptic quadratures.

The characteristic equation for determination of the type of the singularity at the point $Q_1$ reads
\begin{equation*}
\mu^4+d_2\mu+d_0=0,
\end{equation*}
where
\begin{equation*}
\begin{array}{l}
d_2=-2ab[(a-b)^2(h+4\varepsilon_1^2ab)-4\lambda^2ab],\\[5mm]
d_0=8a^3b^3[2\varepsilon_2^2(a-b)^4+\\[5mm]
\varepsilon_1^2(a-b)^2((a-b)^2(h+2\varepsilon_1^2ab)-
4\lambda^2ab)-\lambda^2(h(a-b)^2-2\lambda^2ab)].
\end{array}
\end{equation*}
Note that the negative sign of the value
\begin{equation*}
d_2^2-4d_0=4a^2b^2(a-b)^4(h^2-16\varepsilon_2^2ab)
\end{equation*}
indicates that the singularity of rank $1$ is a focus.

At the point $Q_2$ the solution is
\begin{equation}\label{eq_8}
\begin{array}{l}
M_1=M_2=\alpha_3=\beta_3=0,\\[3mm]
\alpha_1=-a\cos(\varphi),\alpha_2=a\sin(\varphi),\\[3mm]
\beta_1=b\sin(\varphi),\beta_2=b\cos(\varphi).
\end{array}
\end{equation}
\begin{equation*}
(\dot\varphi)^2=4[\varepsilon_1^2(a-b)^2\sin^2\varphi-4\varepsilon_2(a-b)\cos\varphi-2\lambda\varepsilon_1(a-b)\sin\varphi+2h+\lambda^2].
\end{equation*}
This equation can be solved in terms of elliptic quadratures.

The type of the singularity that corresponds to the point $Q_2$ can be determined from
the characteristic equation
\begin{equation*}
\mu^4+d_2^\prime\mu+d_0^\prime=0,
\end{equation*}
where
\begin{equation*}
\begin{array}{l}
d_2^\prime=2ab[(a+b)^2(h-4\varepsilon_1^2ab)+4\lambda^2ab],\\[5mm]
d_0^\prime=-8a^3b^3[2\varepsilon_2^2(a+b)^4+\\[5mm]
\varepsilon_1^2(a+b)^2((a+b)^2(h-2\varepsilon_1^2ab)+
4\lambda^2ab)-\lambda^2(h(a+b)^2+2\lambda^2ab)].
\end{array}
\end{equation*}
Here the value of the expression
\begin{equation*}
{d_2^\prime}^2-4d_0^\prime=4a^2b^2(a+b)^4(h^2+16\varepsilon_2^2ab)
\end{equation*}
is always non-negative.

\section{Conclusion}

The constructed periodic solutions (\ref{eq_5}), (\ref{eq_6}), (\ref{eq_7}) and (\ref{eq_8}) are singularities of rank 1 of the momentum map. The image of these solutions under the momentum map are singular points in the bifurcation diagram. These solutions play a key role in the construction of the atlas of the bifurcation diagram for the
generalized two-field gyrostat. A similar study of the atlas of a bifurcation diagram was carried out by M.\,P.~Kharlamov for the motion of a Kowalevski top in the double field of forces.

\section{Acknowledgments}

The authors are grateful to A.\,V.~Bolsinov for extremely valuable discussions and guidance, National Science Foundation of Poland,
Stefan Banach International Mathematical Center, Warsaw Center of Mathematics and Computer Science,  Faculty of Physics and Astronomy
University of Zielona Gora  and the Organizers of FDIS-2015 especially V.\,S.~Matveev, S.\,L.~Tabachnikov, A.~Maciejewski for hospitality and support.

\bibliographystyle{elsarticle-num}
\bibliography{ryab_bib_newn}

\end{document}